\documentstyle[preprint,aps]{revtex}

\begin{document}

\thispagestyle{empty}
\setcounter{page}{1}

\title{DAMPING RATE OF QUASIPARTICLES IN DEGENERATE
ULTRARELATIVISTIC  PLASMAS}

\author{Michel Le Bellac} 

\address{Institut Non Lin\'eaire de Nice\\
UMR 129 1361 Route des Lucioles\\
06560 Valbonne (FRANCE)\\
and,}

\author{Cristina Manuel}

\address{Dpt. Estructura i Constituents de la Mat\`{e}ria\\
Facultat de F\'{\i}sica,
Universitat de Barcelona \\
Diagonal 647,
08028 Barcelona (SPAIN)} 

\maketitle

\thispagestyle{empty}
\setcounter{page}{0}

\begin{abstract}
$\!\!$
We compute the damping rate of a fermion in a dense relativistic plasma at
zero temperature. Just above the Fermi sea, the damping rate is dominated
by the exchange of soft magnetic photons (or gluons in QCD) and is
proportional to $(E-\mu)$, where E is the fermion energy and $\mu$ the
chemical potential. We also compute the contribution of soft electric
photons and of hard photons. As in the nonrelativistic case, the
contribution of longitudinal photons is proportional to $(E-\mu)^2$, and is
thus non leading in the relativistic case. 
\end{abstract}

\vfill

\hfill\break
\hbox to \hsize{ECM-UB-PF-96/15, \hfill INLN 96/19} 
\hbox to \hsize{September/1996}
\vskip-12pt
\eject

\baselineskip= 20pt
\pagestyle{plain}
\baselineskip= 20pt
\pagestyle{plain}

The properties of quasiparticles in ultrarelativistic (UR)
plasmas have attracted much attention in a recent past \cite{LeBellac}.
A crucial property of a quasiparticle is its decay (or damping)
rate: a quasiparticle which propagates in a plasma is not stable,
as it undergoes collisions with the other particles of the plasma,
and the very concept of a quasiparticle makes sense only if its damping
rate is small enough.

The damping rate of an electron propagating in a nonrelativistic (NR)
plasma was computed almost forty years ago by Quinn
and Ferrell \cite{Quinn}, \cite{Pines}. At first sight, this damping rate is infinite,
due to the singular behavior of the Rutherford cross-section at small
angles. However Quinn and Ferrell were able to obtain a
finite result because the Coulomb interaction is screened in a plasma
(Debye screening), and in the case of a degenerate plasma, they showed that
the damping rate is proportional to $(\varepsilon_p-\varepsilon_F)^2$, where
$\varepsilon_p=p^2/2m$ is the NR kinetic energy
and $\varepsilon_F$ the Fermi energy, when $\varepsilon_p$ is slightly larger
than $\varepsilon_F$. The damping rate remains finite for a non zero
temperature $T$:
only the value of the Debye screening length is modified.

It is interesting to extend the calculation of the damping rate to
the case of relativistic  plasmas: one may have in mind either
electromagnetic (QED) plasmas, such as found in white
dwarves or in the core of nascent neutron stars, or  chromodynamic (QCD)
plasmas such as the quark-gluon
plasma which is
believed to be formed for large enough values of the
temperature $T$ and/or the chemical potential $\mu$.
The NR results are not easily transposed to the relativistic case,
because the exchange of magnetic (or transverse) photons
 in QED or of magnetic gluons in QCD becomes important,
while in the NR case it is suppressed by powers of $(v/c)^2$ with respect to
the exchange of electric  (or longitudinal) gauge
bosons, and is usually neglected. These magnetic photons, or
gluons,
give rise to severe infrared (IR) divergences which are not
easily cured, because there is no static magnetic screening analogous to
Debye screening in the electric case,
but only a weaker
dynamical screening \cite{Weldon}. In many cases, this dynamical
screening is sufficient to remove the IR divergences \cite{BT}, \cite{Baym}
but it has been
known for some time that it cannot solve easily the IR
problem of the damping rate \cite{Pisarski}, at least for non zero $T$.
In a recent
paper \cite{BIancu}, Blaizot and Iancu were nevertheless able to
derive a finite result in the $T\not= 0$, $\mu= 0$ case, by
using a non perturbative approach to resum the leading divergencies.
However they also discovered
that the decay law is no longer exponential.

In this Letter, we address the problem of computing the
damping rate of quasiparticles in degenerate UR plasmas.
 For the sake of definiteness, we treat the case of a
 QED plasma, but our results may be trivially extended to the
QCD case by substituting to the QED coupling $e$ the QCD coupling $g$,
and by taking into account some color group factors. In this computation,
the basic physical idea is that the collisions of the charged quasiparticle
with the particles in the plasma are
governed by photon exchange, and that one must take into
 account the fact that the photon propagator is dressed by the
interactions. Actually, this approach is a particular case of the
resummation method  proposed by Braaten and
Pisarski \cite{BP1}, which relies on the properties of the so-called ``hard thermal
loops" \cite{BP1}, \cite{FT}
 or ``hard dense loops" in the degenerate case \cite{Manuel}. Braaten
and Pisarski pointed out the importance of a
hierarchy of scales, based on the existence of   a ``hard scale" of order
$T$ (or $\mu$), and a ``soft scale" of order $eT$
(or $e\mu$), with $e\ll 1$. When soft scales are involved,
one must use dressed (or resummed) propagators and vertices instead of the
bare ones in a perturbative expansion. An important feature of the
resummation method is that it leads to gauge independent results, due to
the gauge independence of the hard thermal (or dense) loops.

Our main result is that in the case $T=0,\ \mu\not= 0$, dynamical screening
is able to cure the IR divergences of the damping rate due to magnetic
photon exchange in UR plasmas; however, in contrast to the NR case, the
damping rate is dominated by magnetic exchange and is proportional to
$(E-\mu)$, where $E$ is the relativistic energy of the quasiparticle, while
electric photon exchange gives a contribution proportional to $(E-\mu)^2$,
as in the NR case, which may be in fact obtained as a low velocity limit of
the relativistic one \cite{long}. Note, however, that by convention energies and
chemical potentials differ by the rest mass of the particle in the NR and
relativistic cases. Note also that we use a system of units where
$\hbar=c=k_B=1$, and that we follow closely the notations of \cite{LeBellac}.

\begin{figure}
%%Begin InstantTeX Picture
\let\picnaturalsize=N
\def\picsize{4in}
\def\picfilename{resum.eps}
%If you do not have the picture file add:
%\let\nopictures=Y
%to the beginning of the file.
\ifx\nopictures Y\else{\ifx\epsfloaded Y\else\input epsf \fi
\let\epsfloaded=Y
\centerline{\ifx\picnaturalsize N\epsfxsize \picsize\fi \epsfbox{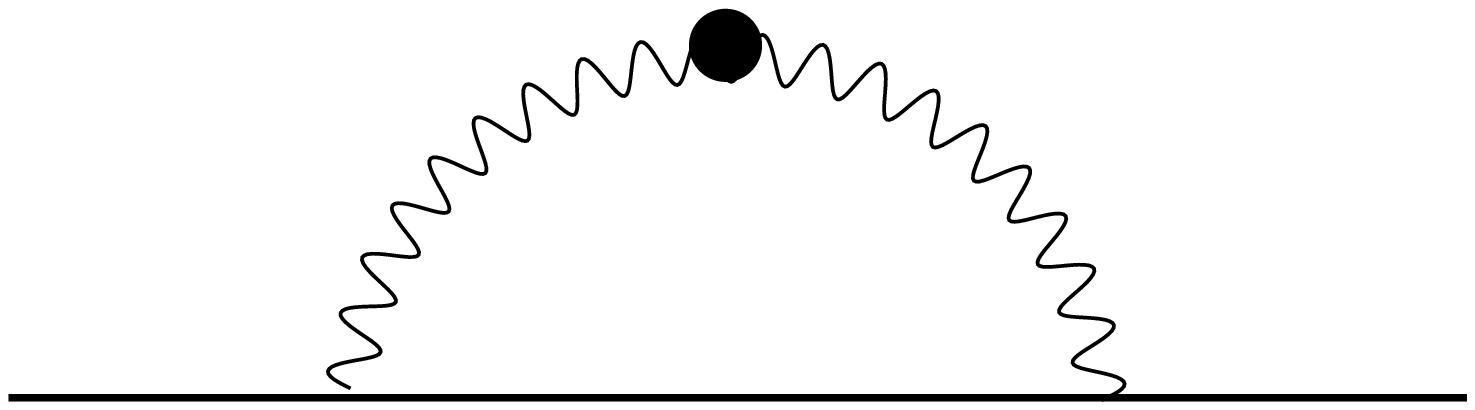}}}\fi
%%End InstantTeX Picture
\vspace*{-2cm}
\caption{Resummed one-loop self-energy of the fermion.}
\label{1loop}
\end{figure}

Let us now proceed to the derivation of our result. We assume that the
quasiparticle energy $E$ is hard (this is automatically ensured in the case
of a degenerate plasma). The damping rate $\gamma(E)$ is given by the
imaginary part of the quasiparticle self-energy $\Sigma$ \cite{Weldon2}; more
precisely
\begin{equation}
 \gamma(E)=-{1\over 4 E}  {\rm Tr}\,\left[{\rm Im}\,\Sigma(p_0+i\eta\,{\bf
p})({P\llap{/\kern1pt}}+m)\right] \Big|_{p_0=E}  \ ,
\label{eqno(1)}
\end{equation}
where $m$ is the electron mass, $E=(p^2+m^2)^{1/2}$, $\eta \rightarrow 0^+$, 
and we have used the by now standard notation: $P_\mu=(p_0,{\bf p})$; 
the lowest
order graph for $\Sigma$ is drawn in Fig. \ref{1loop}. We are mainly interested in
the contribution of soft photons, so that the electron-photon vertex and
the electron propagator may be replaced by the bare ones 
\cite{LeBellac}, \cite{BP1}: only the
photon propagator need be dressed.

We perfom the calculation of $\Sigma$ in the imaginary time formalism; then
the (free) electron propagator is given by
\begin{equation}
S_f(i\omega_n,{\bf k})=\int_{-\infty}^\infty{{\rm d}k_0\over
2\pi}\,{( {K\llap{/\kern3pt}}+m)\rho_f(K)\over
k_0-i\omega_n-\mu} \ , 
\label{eqno(2)}
\end{equation}
with
\begin{equation}
\rho_f(K)=2\pi\varepsilon(k_0)\delta(k_0^2-E_k^2) \ .
\label{eqno(3)}
\end{equation}
In (\ref{eqno(2)}), $\omega_n=\pi(2 n+1)T$ is a fermionic Matsubara frequency and
$\varepsilon(k_0)=k_0/|k_0|$. The (resummed) photon propagator
$\Delta_{\mu\nu}(Q)$ is written in the Coulomb gauge
\begin{equation}
\Delta_{\mu\nu}(Q)=\delta_{\mu 0}\delta_{\nu 0} \, \Delta_L(Q)
+(\delta_{ij}-\hat q_i\hat q_j)\Delta_T(Q) \ ,
\label{eqno(4)}
\end{equation}
where the spectral representations of $\Delta_T$ and $\Delta_L$ read
\begin{mathletters}
\label{eqno(5)}
\begin{eqnarray}
\Delta_L (i \omega_s, q)   & = & 
\int_{-\infty}^{\infty}{{d q_0 \over 2\pi}  \frac{ \rho_L (q_0,q) } {q_0-i \omega_s }}  
- {1 \over q^2 } \ ,  \\
 \Delta_T  (i \omega_s, q) &  =  &
\int_{-\infty}^{\infty}{{d q_0 \over 2\pi}  \frac{ \rho_T (q_0,q) } {q_0-i \omega_s}} 
 \ .
\end{eqnarray}
\end{mathletters}
$\!\!$In (\ref{eqno(4)}) and (\ref{eqno(5)}),
 ${\hat q}_i={\bf q }_i/|{\bf q}|$ and $\omega_s=2\pi s T$ is a
bosonic Matsubara frequency. The explicit expressions of the longitudinal
and transverse spectral functions $\rho_L$ and $\rho_T$ are found by taking
the imaginary parts of $\Delta_L$ and $\Delta_T$
\begin{mathletters}
\label{eqno(6)}
\begin{eqnarray}
\rho_L(q_0,q)=& 2\,{\rm Im}\,\Delta_L(q_0+i\eta,q) &
= 2\,{\rm Im}\,{-1 \over q^2+3 \,\omega_P^2 \left(1-{x\over 2}\ln{\frac{x+1}
{x-1}} \right)} \ ,
\\
\rho_T(q_0,q)=& 2\,{\rm Im}\,\Delta_T(q_0+i\eta,q) &
= 2\,{\rm Im}\,{-1\over (q_0+i\eta)^2-q^2-{3\over
2}\,\omega_P^2\left(x^2+{x(1-x^2)\over 2}\ln{x+1\over
x-1}\right)} \ ,
\end{eqnarray}
\end{mathletters}
$\!\!$where $x=(q_0+i\eta)/q$, $\omega_P=M/\sqrt 3$ is the plasma frequency which is
related to the Debye mass $M$ given by
\begin{equation}
M^2={e^2 \over \pi^2}\left(\mu^2+{\pi^2T^2\over 3}\right) \ .
\label{eqno(7)}
\end{equation}

The diagram in Fig. \ref{1loop} is now evaluated in the imaginary time formalism
$(P=(i\omega_n, {\bf p}))$
\begin{equation}
\Sigma(P)= e^2T\sum_s\int{{\rm d}^3q\over (2\pi)^3}\gamma_\mu \,
S_f(i(\omega_n-\omega_s),{\bf p-q})\gamma_\nu \,
\Delta_{\mu\nu}(i\omega_s,{\bf
q}) \ .
\label{eqno(8)}
\end{equation}

The sum over Matsubara frequencies is easily performed when one plugs in
(\ref{eqno(8)})  the spectral representations (\ref{eqno(2)})  and 
(\ref{eqno(6)}) of the propagators. Taking the
imaginary part of $\Sigma$ after the analytical continuation $i\omega_n+\mu\to
p_0+i\eta$ to Minkowski space, and taking the trace in (\ref{eqno(1)}), one finds for
the damping rate, with ${\bf k} = {\bf p} - {\bf q}$,
\begin{eqnarray}
\label{eqno(9)}
\gamma(E) & =&{\pi e^2\over E} \int{{\rm d}^3q\over
(2\pi)^3}\int_{-\infty}^\infty{{\rm d}k_0\over 2\pi}\rho_f(k_0)
\int_{-\infty}^\infty{{\rm d}q_0\over 2\pi} \\
&\times &
\left(1+ n(q_0)- {\tilde n}(k_0)\right)  \delta(E-k_0-q_0)
\left\{[p_0k_0+{\bf p\cdot
k}+m^2] \right. \nonumber \\
&\times & \left. \rho_L(q_0,q)
+2[p_0k_0-({\bf p\cdot \hat q})({\bf k\cdot\hat
q})-m^2]\rho_T(q_0,q)\right\}  \ .
\nonumber
\end{eqnarray}
In (\ref{eqno(9)}), $n$ and ${\tilde n}$ are Bose-Einstein and Fermi-Dirac
distribution functions ($\beta=1/T$)
\begin{equation}
n(q_0)={1\over {\rm e}^{\beta q_0}-1} \ , \qquad
{\tilde n} (k_0)={1\over {\rm e}^{\beta (k_0-\mu)}+1} \ .
\label{eqno(10)}
\end{equation}

Eq. (\ref{eqno(9)}) could have also been derived from kinetic theory
\cite{Luttinger}, using standard
identities between the distribution functions (\ref{eqno(10)}).
The case $k_0<0$ (see (\ref{eqno(3)}))
corresponds in kinetic theory to $e^{+}-e^-$ annihilation, which is not
IR singular and is even absent in the $T=0$ case. We thus concentrate on
the $k_0>0$ case, which corresponds in kinetic theory to $e^{-}-e^-$
scattering. From now on, we shall also restrict ourselves to the $m=0$ and
$T=0$ case, leaving the general case to a forthcoming publication by one of
the authors \cite{long}. In the $T=0$ limit, $(1+n(q_0))=\Theta(q_0)$ and
${\tilde  n}(k_0)=\Theta(\mu-E+q_0)$, where $\Theta$ is the step function.
The  $q_0$ integration is then limited by
\begin{equation}
0\leq q_0\leq E-\mu \ .
\label{eqno(11)}
\end{equation}
This is also easily seen in kinetic theory, since, due to Pauli blocking,
the quasiparticle can only scatter into states with energy $E_{|{\bf
p-q}|}$ such that $E_{|{\bf p-q}|}\leq E$ and furthermore the particle
on which it scatters must leave the Fermi sea, so that 
$ E_{|{\bf p-q}|}\geq \mu$.
 Note also that the exchanged photon must be space-like: $Q^2<0$, so
that the pole part of $\rho_{L,T}$ \cite{LeBellac} does not contribute.

Now, the IR singular contribution comes from small values of the photon
momentum $q$; in order to isolate this kinematical region, we follow
Braaten ans Yuan \cite{BY} and introduce an intermediate cut-off $q^*$ such that
$e\mu \ll q^*\ll \mu$. The ``soft" region is defined by $q<q^*$, the ``hard"
one by $q>q^*$: in this latter region we may take the $M^2=0$ limit in the
denominators of the spectral functions
$\rho_{L,T}$ in (\ref{eqno(9)}) \cite{LeBellac}. Let us concentrate on 
the soft region, where we can make the approximation
\begin{equation}
E_{|{\bf p-q}|}=E-q_0\simeq E-{\bf \hat p\cdot q} \ .
\label{eqno(12)}
\end{equation}

Keeping only the leading terms in (\ref{eqno(9)}), we find the contribution from the
soft region to $\gamma(E)$
\begin{eqnarray}
\label{eqno(13)}
\gamma_{\rm soft}(E) & \simeq & {e^2\over 2} \int {{\rm d}^3q \over
(2\pi)^3} \left( \Theta(q_0) \, -\Theta(\mu-E+q_0) \right) \, \Theta(q^*-q) 
\nonumber \\
&  \times & \left\{ \rho_L(q_0, q)+(1-\cos^2\theta)\rho_T(q_0,q) \right\}
 \ ,
\end{eqnarray}
with $q_0={\bf \hat p\cdot q}=q\cos \theta$. It is convenient to use as
integration variables $q_0$ and $q$, the integration domain $D$ being
\begin{equation}
D:\ \{0\leq q_0\leq E-\mu;\ q_0\leq q\leq q^*\} \ .
\label{eqno(14)}
\end{equation}
Then (\ref{eqno(13)}) becomes $(x=q_0/q$)
\begin{eqnarray}
\gamma_{soft} (E)  & \simeq & {e^2 M^2  \over 4 \pi}   
 \int_{D}  { dq_0 \, dq } 
 \left\{ 
 \frac{q_0} { 2 \left[  q^2 + M^2 Q_1(x )  \right]^2  +
 {M^4 \pi^2 x^2 \over 2 } }
 \right. 
\nonumber \\
& + &
\left. \frac {q_0} {  \left[ 2\, q^2 + M^2 Q_2(x)  \right]^2  +
 {M^4 \pi^2 x^2 \over 4 } }
 \right\}  \ ,  
\label{eqno(15)}
\end{eqnarray}
 where  
\begin{equation}
\label{Legendre}
Q_1 (x)  =  1 -  \frac{x }{2}
\ln{\frac{1+x} {1- x} } \ ,
\qquad 
Q_2 (x)  =  - Q_1 (x) + {1 \over 1 - x^2  } \ . 
\end{equation}

Note that in the absence of screening (namely, by setting $M=0$ in the
denominators of (\ref{eqno(15)})), one would get IR divergent integrals. In general,
the integrals in (\ref{eqno(15)}) must be computed numerically. Fortunately, it is
possible to derive an accurate analytical result in the physically
interesting case $(E-\mu)\ll M$. Indeed, it is easy to check that in this
region one may expand the denominators in (\ref{eqno(15)}) in powers
of $q_0$. Keeping only the leading terms, the first denominator
 in (\ref{eqno(15)}), corresponding to longitudinal photon
exchange, may be replaced by $(q^2+M^2)$, which leads to Debye screening.
The second denominator in (\ref{eqno(15)}), corresponding to transverse photon
exchange, may be replaced by
\begin{equation}
4\,q^4+{\pi^2M^4x^2\over 4} + 8 M^2 q^2 x^2 \ .
\label{eqno(16)}
\end{equation}
It can be shown that the last term in (\ref{eqno(16)}) gives a subdominant 
contribution \cite{long},  while the second term leads to the usual form of
dynamical screening \cite{Weldon} - \cite{Baym}. Computing separately the
longitudinal and transverse contributions, we find, with $u^* =( q^*/ M)^2$,
\begin{eqnarray}
\label{eqno(17)}
\gamma_{soft }^{L} (E) &  \simeq & {e^2 (E-\mu)^2  \over 32 \pi M} 
\int_{0}^{u^*} {  du  \over \sqrt{u} (u+1)^2 }   \simeq 
{e^2 M^2  \over 16 \pi}\,   
 (E-\mu)^2 \left( \frac{\pi}{4 M^3} - \frac{1}{3 q^{* 3}} \right)        \ ,
\\
\label{eqno(18)}
\gamma_{soft }^{T} (E) &  \simeq & {e^2 M  \over 4  \pi^3 } 
\int_{0}^{u^*}   du  \, \sqrt{u} \, 
 \ln {\left(  1 +  \frac{ \pi^2 (E- \mu)^2}{ 16 \, M^2\, u^3 }  \right)}  \nonumber
\\ &  \simeq & 
 {e^2  \over 24  \pi }\, (E-\mu) +  
 {e^2 M^2  \over 32 \pi }\,  
 (E-\mu)^2 \left ( - \frac{1}{3 q^{* 3}} \right)   
 \ ,
\end{eqnarray}
where we have only kept the leading terms in $(E-\mu)$ and $1/q^*$.

The total contribution of the soft region to the decay rate is obtained by
adding the longitudinal and transverse contributions to get
\begin{equation}
\gamma_{soft } (E)   \simeq {e^2  \over 24  \pi }\, (E-\mu) +
{e^2 M^2  \over 32 \pi}\,   
 (E-\mu)^2 \left( \frac{\pi}{ 2 M^3} - \frac{1}{ q^{*3}} \right)      \ .
\label{eqno(19)}
\end{equation}
The transverse contribution dominates over the longitudinal one for small
values of $(E-\mu)$.

We finally evaluate the contribution from the hard region.
Since we are only interested in extracting 
 the leading dependence in the fermionic energy of the decay rate
we will use a simple approach to compute the hard contribution. 
It is possible to 
recover bare or unresummed perturbation theory to order $e^4$ by using 
the spectral densities (\ref{eqno(6)}) neglecting $M^2$ in 
the denominators \cite{LeBellac}.
This is only valid in the momentum transfer region $q > q^*$. Therefore
one finds for the hard contribution to the decay rate
\begin{equation}
\label{eqno(20)}
\gamma_{hard} (E)   \simeq  {e^2 M^2  \over 8 \pi }    
\int^{q_m}_{q*} {  dq } \int_{0}^{E-\mu} d q_0 \left\{ 
 \frac{q_0} {  q^4  }
 +  \frac{q_0} {   2 \, q^4  }
 \right\}  \ . 
\end{equation}
After an straightforward computation one finds

\begin{equation}
\label{eqno(21)}
\gamma_{hard} (E)  \simeq   {e^2 M^2  \over 32 \pi}\,   
 (E-\mu)^2 \left(   \frac{1}{ q^{* 3} } - \frac{1}{ q_{max} ^3} \right)  \ ,
\end{equation}
where $q_{max} \simeq \mu$ is the maximum momentum transfer that it is allowed
by kinematics.

The total  decay rate is found just by adding the
soft and hard contributions. Then one finds that the dependence 
on the scale $q^*$ cancels, as it should. The result is 
\begin{equation}
\label{eqno(22)}
\gamma (E)   \simeq  {e^2  \over 24 \pi }\, (E-\mu) +
 {e^2 M^2  \over 32 \pi}\,   
 (E-\mu)^2 \left(   \frac{\pi}{2 M^3} - \frac{1}{ q_{max} ^3} \right)    
  \ .
\end{equation}

In conclusion, we have been able to compute the damping rate of a
quasiparticle in a degenerate ultrarelativistic plasma, when the fermion
energy $E$ is just above the Fermi energy $\mu$. This damping rate is
dominated by transverse photon (or gluon) exchange and proportional
to $(E - \mu)$. This behavior arises from the combined effect of 
dynamical screening and phase space restrictions due to Pauli blocking.
The lifetime $\tau$ is related to $\gamma$ by $\tau \sim 1/\gamma$, and
therefore the lifetime becomes infinite as the fermion energy approaches
the Fermi energy, so that the Fermi sea is stable.	

\vfill
\eject
{\bf Acknowledgments:}

After this work was completed, we learned that our results have been
obtained independently by J.-Y.~Ollitrault and B.~Vanderheyden.

C.~M. wants to thank C.~Lucchesi,
N.~Rius, A.~Ramos, and  J.~Soto  for useful discussions. She
is also thankful to the Institut Non Lin\'eaire de
Nice, and to the Institut de Physique de l' Universit\'e de 
Neuch\^atel for hospitality. 

This work has been supported in part by funds provided by the
European Contract CHRX-CT93-0357, ``Physics at High Energy
Accelerators", by the Swiss National Science Foundation, and
by the Ministerio de Educaci\'on y Ciencia (Spain).

\end{document}